\documentclass[aip,jcp]{revtex4-1}

\usepackage{graphicx}
\usepackage{dcolumn}% Align table columns on decimal point
\usepackage{bm}% bold math
\usepackage{hyperref} % Needed for the table
\usepackage{multirow} % Needed for the table
\usepackage{soul}
\newcommand{\beq}{\begin{equation}}
\newcommand{\eeq}{\end{equation}}
\newcommand{\bea}{\begin{eqnarray}}
\newcommand{\eea}{\end{eqnarray}}
\usepackage{wrapfig}

\begin{document}

\title{Effects of non-pairwise repulsion on nanoparticle assembly}

\author{Sawyer S. Hopkins, Amitabha Chakrabarti, and Jeremy D.\ Schmit\footnote{schmit@phys.ksu.edu}}
\affiliation{Department of Physics, Kansas State University,
 Manhattan, KS 66506, USA }

\begin{abstract}
Electrostatic interactions provide a convenient way to modulate
interactions between nanoparticles, colloids, and biomolecules
because they can be adjusted by the solution pH or salt
concentration. While the presence of salt provides an easy method to
control the net interparticle interaction, the nonlinearities
arising from electrostatic screening make it difficult to quantify
the strength of the interaction. In particular, when charged
particles assemble into clusters or aggregates, nonlinear effects
render the interactions strongly non-pairwise. Here we report
Brownian dynamics simulations to investigate the effect that the
non-pairwise nature of electrostatic interactions has on
nanoparticle assembly. We compare these simulations to a system in
which the electrostatics are modeled by a strictly pairwise Yukawa
potential. We find that both systems show a narrow range in
parameter space where the particles form well-ordered crystals.
Bordering this range are regions where the net interactions are too
weak to stabilize aggregated structures, or strong enough that the
system becomes kinetically trapped in a gel. The non-pairwise
potential differs from the pairwise system in the appearance of an
amorphous phase for strongly charged particles. This phase appears
because the many-body electrostatic interactions limit the maximum
density achievable in an assembly.
\end{abstract}

\maketitle % this produces the title block

\section{Introduction}
The self-assembly of particles into ordered structures requires a
tuning of the net interparticle potential such that the interactions
are sufficiently strong to stabilize the structure, yet thermally
reversible so that defects can be removed
\cite{Rapaport2008,Jack2007,Whitelam2015a}. Electrostatic
interactions are a convenient mechanism for this tuning since they
can be adjusted by modifying the surface chemistry, solution pH, or
salt concentration \cite{Stradner2005}. Electrostatic tuning played
a central role in early descriptions of colloid stability where
electrostatic repulsion was used as a balance against fixed
attractive interactions \cite{Derjaguin1941,Verwey1948}. The DLVO
framework has remained an important guide for the solubility of
charged particles, however, important differences emerge when
considering particles on the nanometer scale such as biomolecules or
nanoparticles. For starters, the long-ranged van der Waals
attraction becomes less important than short-ranged interactions
like H-bonds and the hydrophobic effect. This has inspired numerous
studies of the competition between short-range attraction and
long-range repulsion, where the repulsion is modeled using a
repulsive Yukawa potential
\cite{Sciortino2004,Mossa2004,Archer2007a,Bomont2010,Lee2010a,Bollinger2016,Zhuang2016}.
This qualitatively captures the effects of Coulomb repulsion
screened by salt because the Yukawa potential emerges from the small
potential limit of the Poisson-Boltzmann (PB) equation. However, as
the density of charges increases, such as when the particles
aggregate, the electrostatic potential rises above the acceptable
range for the small potential (Debye-Huckel) treatment. While this
has the expected effect of causing quantitative discrepancies as the
nonlinearities of the PB equation take effect, a less intuitive
consequence is that aggregation qualitatively changes the nature of
the interparticle interaction. This is because aggregation
compresses the screening layer around each particle resulting in a
{\em favorable} Coulomb interaction, but incurring a large entropic
penalty \cite{Schmit2011gel}.

An important consequence of the transition to entropy-dominated
electrostatics is that electrostatic repulsion can no longer be
treated in a pairwise fashion (i.e. screened Coulomb interactions).
Instead, the electrostatic interactions are delocalized through the
screening layers. From a theoretical viewpoint, this has mixed
effects. On one hand, delocalization of the interaction means the
electrostatic free energy is insensitive to the precise location of
charges, which makes the calculation amenable to mean-field
treatments \cite{Schmit2010,Dahal2018}. On the other hand, it means
that the calculation must account for the non-pairwise nature of the
interactions.

The purpose of this paper is to explore how non-pairwise repulsion
contributes to nanoparticle assembly and aggregation. From a
thermodynamic point of view, we expect that non-pairwise repulsion
will destabilize densely packed structures since each additional
inter-particle contact reduces the affinity of all previous
contacts. Usually, when an aggregation process terminates with the
formation of a gel or amorphous aggregate, it is assumed that this
state is a kinetic trap and the thermodynamic ground state is an
ordered crystal that maximizes the favorable contact energy. In the
presence of non-pairwise repulsive interactions this is not
necessarily the case because the nonlinear repulsion can destabilize
the higher density structure. Note that a similar effect can also
emerge from pairwise interactions if the range of the repulsion is
long enough to permit next-nearest-neighbor interactions. However,
non-pairwise additivity means that there can be a limit on the
packing density even if the screening length is smaller than the
particle size.

Non-pairwise interactions are also expected to have a significant
effect on the kinetics of assembly, because particles attempting to
bind to a cluster will be less likely to explore higher density
states. This will provide a bias against ordered crystals even if
these states are the thermodynamic minimum. The bias against dense
states will be particularly significant in the nucleation phase,
especially if the critical nucleus is small enough that bulk-like
interactions have not yet emerged (as is the case in protein
crystals \cite{Galkin1999}).

In a previous paper we explored the effect of electrostatic
interactions in the competition between crystallization and gelation
\cite{Schmit2011gel}. That work used the simplified criteria that a
gel would emerge when a single interparticle contact provides enough
binding energy to pay the translational entropy cost of removing a
monomer from solution. In that model, the window of successful
crystallization conditions is bounded by the stability of the
crystal and the instability of the solution with respect to two-body
interactions. This window expands when the particle charge and salt
concentration are simultaneously increased such that the net
interaction strength is maintained but the repulsive interactions
are more nearly pairwise \cite{Schmit2011gel}. Here, we report
Brownian dynamics simulations that qualitatively confirm these
predictions with a crystallization window that grows narrower as the
non-pairwise character of the interactions increases. However, our
simulations also show the emergence of an amorphous structure that
is not present when the interactions are strictly pairwise,
indicating that non-pairwise interactions have destabilized the high
density states required for crystallization.

\section{Methods}

\subsection{Brownian dynamics simulations model a system with a hard core, short range attraction, and electrostatic repulsion}

Our simulations consist of a system of particles evolving according
to the Langevin equation.~\cite{Chen2004a}
\begin{equation}
\label{eqlangevin} m_i\ddot{r}_i=F_i(r)-{\Gamma}\dot{r}_i+\xi_i(t)
\end{equation}
where $F_i$ is the systematic force, $\xi_i$ is the stochastic
force, $m_i$ is the particle mass, and $\Gamma$ is the friction
coefficient. The systematic force can be subdivided into
electrostatic and non-electrostatic contributions. The
non-electrostatic forces consist of a short ranged attractive
interaction with a hard core repulsion, $F_a\left(r\right)$, while
the electrostatic part, $F_{es}\left(r\right)$, is repulsive at all
distances. The short range attractive force encompasses H-Bond, van
der Waals, and hydrophobic effects, and was modeled using an
extended Lennard-Jones (LJ) potential of the form shown in equation
\ref{eqlj}, in which the particle diameter $\sigma = 1$, $k_BT = 1$,
and $\epsilon$ is used as a control variable to tune the attractive
strength of the force.

\begin{equation}
\label{eqlj} V_a\left(r\right) = -\int F_a\left(r\right) \mathrm{d}r
=
4{\epsilon}k_BT\left[\left(\frac{\sigma}{r}\right)^{50}-\left(\frac{\sigma}{r}\right)^{25}\right]
\end{equation}

The simulations consisted of 2500 uniform spherical particles in a
square box at a volume concentration of 9.5\%. The particles evolved
through the integration of Eq. \ref{eqlangevin} with a time-step of
0.001 (unit-less). Periodic boundary conditions were implemented
across all surfaces. Integration was halted for all trials after
$10^8$ integration cycles. The attractive potential strength,
$\epsilon$ was varied in $0.25K_BT$ increments. To assist in the
initial nucleation of crystalline structures, a 4x4x4 primitive
cubic seed crystal was placed in the center of the box at the
beginning of each simulation.

\subsection{Electrostatic repulsion is computed from the volume accessible to the screening layer}
Here we derive an effective potential intended to qualitatively
capture the effects of screening layer distortion on the
electrostatic free energy. The full electrostatic free energy is the
sum of the Coulomb energy and the entropy of the salt
$F_{ES}=E_\mathrm{Coul}-TS_\mathrm{salt}$. These terms are given by
\cite{Overbeek1990,Sharp1990,Andelman1995}
\begin{eqnarray}
E_{Coul}& =& \frac{1}{2 \epsilon}\int_V |\nabla \Psi\left(r\right)|^2 \mathrm{d}V \label{eq_Coulomb}\\
-TS_\mathrm{salt} &=& k_B T\int_V
\left[c_+ln\left(\frac{c_+}{c_s}\right)- c_+ + c_s \right] +
\left[c_-ln\left(\frac{c_-}{c_s}\right) -c_- + c_s\right]
\mathrm{d}^3r. \label{eq_entropy}
\end{eqnarray}
where $\Psi$ is the electrostatic potential, $c_\pm$ are the local
concentration of cations and anions, $\epsilon$ is the local
permittivity, and $c_s$ is the salt concentration in a reservoir far
from any charges. For monovalent salt, minimization of $F_{ES}$ with
respect to the ion concentrations yields
\begin{equation}
c_\pm=c_s  e^{\mp e\Psi/k_BT}
\end{equation}
These concentrations can be plugged into the Poisson equation to
yield the well-known PB equation for the electrostatic potential.
Rather than take the computationally demanding step of solving the
PB equation at each time step in the simulation, we employ a series
of approximations that allow for a closed form free energy
expression while retaining the essential nonlinearities.

The first approximation is to neglect the Coulomb energy term, which
is a minor contribution to the free energy in the aggregated states
that are of primary interest here \cite{Schmit2010,Schmit2011gel}.
The neglect of the Coulomb energy will have the effect of
artificially narrowing the crystallization window since this term is
attractive for dense structures and repulsive for more diffuse ones
\cite{Schmit2011gel}.

The next approximation is to model the screening layer with a step
function profile. That is, we approximate the potential to have a
constant value $\bar{\Psi}$ within a distance $a$ of the particle
surface and $\Psi=0$ outside this distance \cite{Schmit2018}. The
potential within the screening layer can be determined from the
condition that the layer contains enough charge to neutralize the
particle \cite{Schmit2018}
\begin{equation}
\label{eq_neutral} Q = -v\left(c_+ - c_-\right) = 2vc_s\sinh(\frac{e
\bar{\Psi}}{k_BT})
\end{equation}
where $v$ is the volume accessible to the ions in the screening
layer.  With the constant potential approximation, the integrals in
Eq. \ref{eq_entropy} are easily evaluated and we find that the
electrostatic free energy is
\begin{equation}
\label{eq_f_main} \frac{F_{ES}}{k_BT} \simeq -S_\mathrm{salt}/k_B =
Q\left(\mathrm{sinh}^{-1}\left(\frac{1}{\zeta}\right) -
\sqrt{1+\zeta^2} + \zeta \right)
\end{equation}
where $\zeta = {2vc_s}/{Q}$. The electrostatic free energy
associated with a particular macroion depends on the volume $v$
accessible to that ion's screening layer. This volume, in turn, will
depend on the location of neighboring particles, which can occupy
volume that would be otherwise accessible.

\begin{figure}[htb]%t=top, b=bottom, h=here

    \centering
    \includegraphics[height=1.5in]{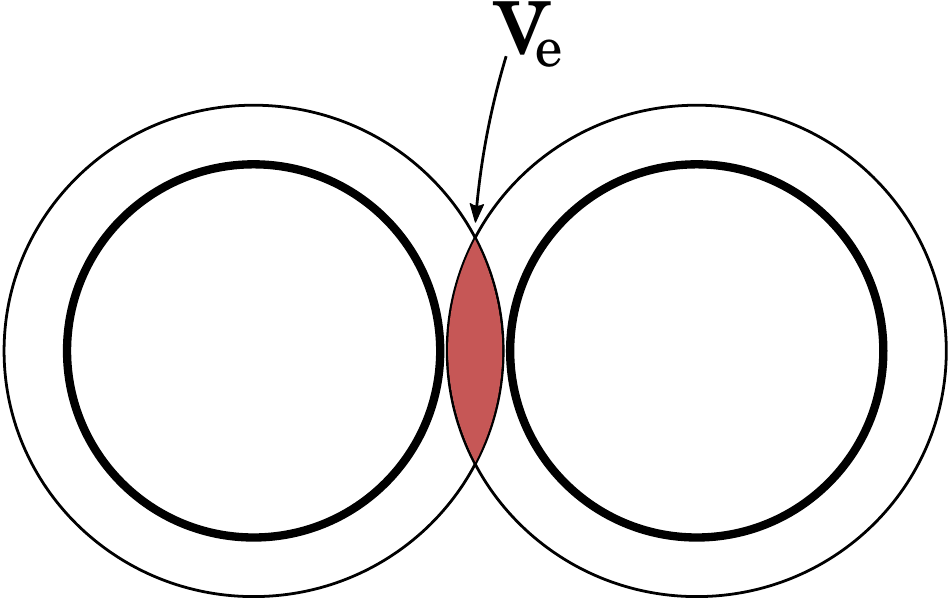}

    \caption[The excluded volume $v_e$ created via the overlap of electrostatic screening layers.]{The excluded volume $v_e$ created via the overlap of electrostatic screening layers.}

    \label{fig_excluded}
\end{figure}

The restriction of the screening layer volume is shown schematically
in  Fig. \ref{fig_excluded}. For the illustrated two-body
interaction, the volume accessible to each screening layer is
\begin{equation}
\label{eq_volume_full} v = v_t - v_p - \frac{1}{2}v_e
\end{equation}
where $v_t$ is the total volume inside a sphere of radius $R +a$,
$v_p=4\pi R^3/3$ is the volume of the macroion, $v_e$ is the volume
of the overlapped region, $R$ is the particle radius, and $a$ is the
screening layer thickness (see Fig. \ref{fig_excluded}). To simplify
the calculation of the screening layer volumes we conduct our
simulations at a salt concentration where the screening layer
thickness is $\sim 10\%$ of the particle radius, since for
$a/R<0.15$ it is impossible for the layers of three different
macroions to overlap in the same region of space. This allows the
overlapped volume to be calculated using pairwise interactions. For
a protein $i$ with $n$ neighbors with centers within a sphere of
radius $2(R+a)$ from the center of $i$, the excluded volume is given
by~\cite{Lekkerkerker2011}
\begin{equation}
 \label{eq_excluded}
v_{e_i}=\sum_{j=1}^n \frac{4\pi}{3}(R+a)^3\left[1 -
\frac{3r_{ij}}{4(R+a)} +
\frac{1}{16}\left(\frac{r_{ij}}{(R+a)}\right)^3\right]
\end{equation}
Note that even though the overlapped volume is calculated based on
pairwise contacts, the free energy is still non-pairwise due to the
nonlinearity of Eq. \ref{eq_f_main}.

To see the effects of the non-pairwise potential, we run an
equivalent set of simulations where the repulsion is described by a
strictly pairwise Debye-Huckel potential
\begin{equation}
\label{eq_big_boy_dh} U_{dh}\left(r\right) =
\gamma\left(Q,c_s\right)a\frac{e^{-r/a}}{r}
\end{equation}
Here $\gamma$ is a renormalized charge that ensures that the
pairwise and non-pairwise potentials yield equivalent interaction
energies for two isolated particles in contact. This is necessary
because the approximations leading to the derivations of both Eq.
\ref{eq_f_main} and the Debye-Huckel potential lead to
non-equivalent interactions at the same change and salt
concentrations. The renormalized charge is given by
\begin{equation}
\label{eq_pair_wise} \gamma =
k_BT\frac{R}{a}e^{R/a}F_{ES}\left(v_2,Q,c_s\right)
\end{equation}
where $F_{ES}\left(v_2,Q,c_s\right)$ is evaluated from Eq.
\ref{eq_f_main} using the two-body screening layer volume for two
particles in contact ($r=2R$). In this case the overlap volume
reduces to
\begin{equation}
\label{eq_simple_excluded} v_2 = {\pi}a^2\left[\frac{4a}{3}+2
R\right]
\end{equation}

\subsection{Distinct phases are identified using order parameters sensitive to density and local structure.}

We characterize the aggregated structure using two order parameters.
The first of these is the coordination number, which we calculate by
counting the number of particles with a center-to-center distance
less than $2(R+a)$ from the reference particle.

While the coordination number provides useful information about the
density, we require another metric to probe the structure of an
aggregate. One method to do this is to project the nearest-neighbor
vectors onto spherical harmonics \cite{Steinhardt1983,Lechner2008}
%The second parameter is intended to assess local ordering within the
%aggregate. This order parameter for particle $i$ with $n$ neighbors
%within $2(R+a)$ is calculated by a weighted projection of the
%spherical harmonics of the $n$ neighbors in respect to particle $i$
%~\cite{Steinhardt1983,Lechner2008}.

\begin{equation}
\label{eq_bond_stuff} q_{lm}\left(i\right) =
\frac{1}{n\left(i\right)}\sum_{j=1}^{n\left(i\right)}Y_{lm}\left(r_{ij}\right)
\end{equation}
Here the sum is over the $n(i)$ particles within a center-to-center
distance $2(R+a)$ from a particle $i$. A rotationally invariant
version can be constructed as follows \cite{Steinhardt1983}
\begin{equation}
q_l\left(i\right) =
\sqrt{\frac{4\pi}{2l+1}\sum_{m=-l}^l|q_{lm}\left(i\right)|^2}
\label{eq_ql}
\end{equation}
The ability of this metric to resolve distinct phases improves by
averaging over the first coordination shell \cite{Lechner2008}
\begin{equation}
\bar{q}_{lm}\left(i\right) =
\frac{1}{n\left(i\right)}\sum_{k=0}^{n\left(i\right)}q_{lm}\left(k\right)
\end{equation}
which can also be cast in a rotationally invariant form
\begin{equation}
\label{eq_more_bond_stuff} \bar{q}_l\left(i\right) =
\sqrt{\frac{4\pi}{2l+1}\sum_{m=-l}^l|\bar{q}_{lm}\left(i\right)|^2}
\end{equation}
Of the parameters defined by Eq. \ref{eq_more_bond_stuff},
$\bar{q}_4$ and $\bar{q}_6$ are particularly useful for their
ability to resolve common crystal structures like BCC, FCP, and HCP
\cite{Lechner2008}.

\section{Results and Discussion}

\subsection{Particles interacting by pairwise potentials form liquid, crystal and gel phases}

\begin{figure}[!ht]
\centering
\includegraphics[width=7cm,height=4.6cm]{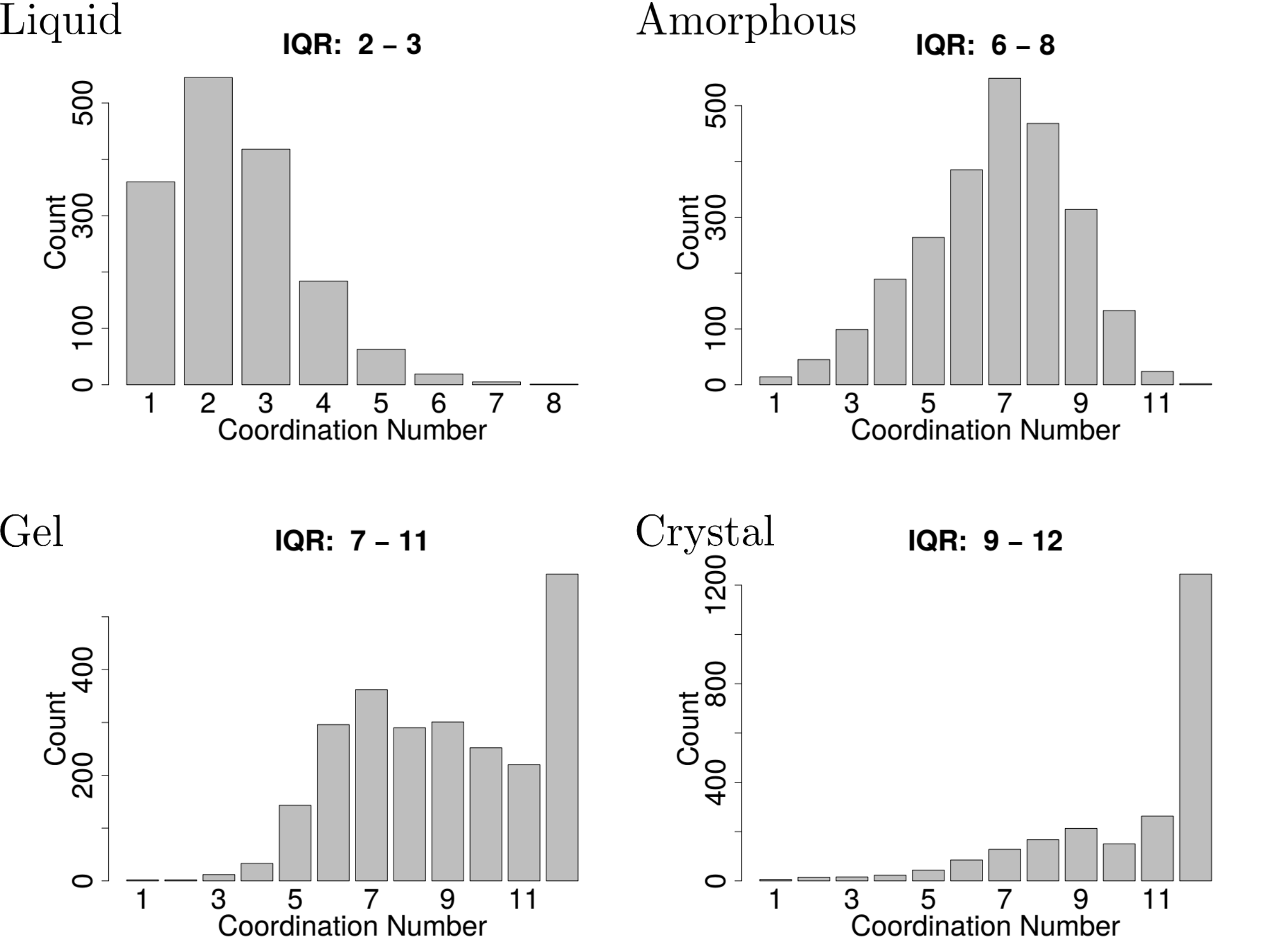}
\caption{Representative coordination number distributions for the
four phases identified in the simulations. The liquid phase is
characterized by transient contacts so most particles have only 1-3
nearest neighbors. The amorphous phase shows a broad range of
coordination numbers, but a conspicuous lack of particles with 11 or
12 nearest neighbors. The gel phase also has a broad distribution of
coordination numbers along with a second peak at 12 nearest
neighbors. This is consistent with HCP and/or FCC packing with a
large surface area. However, the inner quartile range (IQR) of
particles does not include 12. The crystal phase is dominated by
12-fold coordinated particles, indicated a highly ordered structure
and low surface area.} \label{fig_phase_hist}
\end{figure}

Initial analysis of the simulations was conducted by examining the
distribution of nearest neighbor coordination numbers at the end of
the simulation. Representative histograms are shown in Fig.
\ref{fig_phase_hist}. For the simulations interacting via a pairwise
potential we identified three distinct phases from these
distributions.

The liquid phase was defined by coordination histograms where the
peak was three or less. This was observed in simulations where the
particles are strongly charged and/or have a weak LJ binding energy.
Under these conditions the net interparticle attraction is not
sufficient to pay the entropic cost of confining the particles
within an aggregated structure, and the contacts we observe are
transient collisions. Within the liquid regime, the crystal seeds
placed at the start of the simulation dissolve and do not re-form
(data not shown).

\begin{figure}[!ht]
\centering
\includegraphics[width=0.95\linewidth]{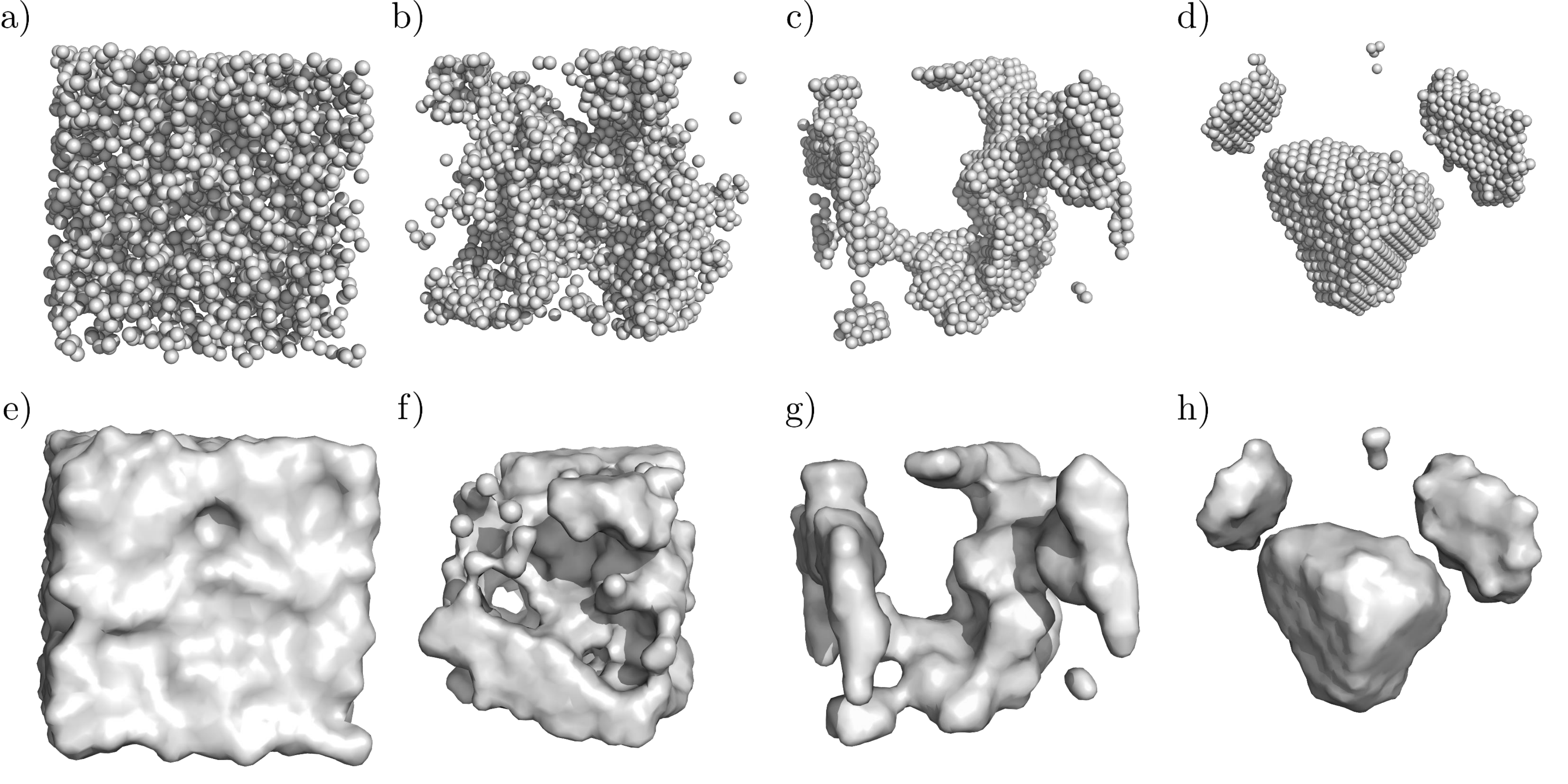}
\caption{The structure of the four characterized phases shown in
both a per particle view (a-d) and a surface overlay (e-h). Solution
(a,e). Amorphous (b,f). Gel (c,g). Crystal (d,h). Periodic boundary
conditions apply to all images.} \label{fig_pymol_all}
\end{figure}

The crystal phase is distinguishable by the sharp peak in the
coordination number histogram at 12. We classify the system in this
phase when the inner quartile range (IQR) of coordination numbers
contains this peak. The crystal phase is observed when the LJ
attraction and electrostatic repulsion combine to give intermediate
values for the net interparticle attraction. These values are strong
enough that six bonds are sufficient to surmount the entropic
penalty for capturing a particle, yet weak enough that the
nucleation of new aggregates is slow on the simulation timescale.
This means that the growth is localized to a small number of
dominant clusters (see Fig. \ref{fig_pymol_all}).

The gel phase also shows a peak in the coordination number
distribution at 12, but differs from the crystal phase in the
greater weight of the distribution at smaller numbers. We classify
the system in the gel phase when the IQR does not contain 12. Visual
inspection (Fig. \ref{fig_pymol_all}), and the abundance of 12-fold
coordinated particles, indicate that the gel has a local structure
similar to the crystal. The primary difference is the much greater
surface area, leading to a structure that spans the simulation box
(Fig. \ref{fig_pymol_all}). This suggests that the gel is metastable
with respect to the crystal phase and that, with enough time, the
system would minimize its surface energy by ripening into a more
compact structure. The formation of a metastable phase is the
expected outcome at high binding energies when nucleation is fast on
simulation timescales and the particle detachment events required
for ripening are slow \cite{Jack2007,Rapaport2008,Whitelam2009a}.

\subsection{Non-pairwise systems have a fourth phase with an intermediate density}

\begin{figure}[!ht]
%\begin{subfigure}{.45\textwidth}
%  \centering
  \includegraphics[width=0.45\linewidth]{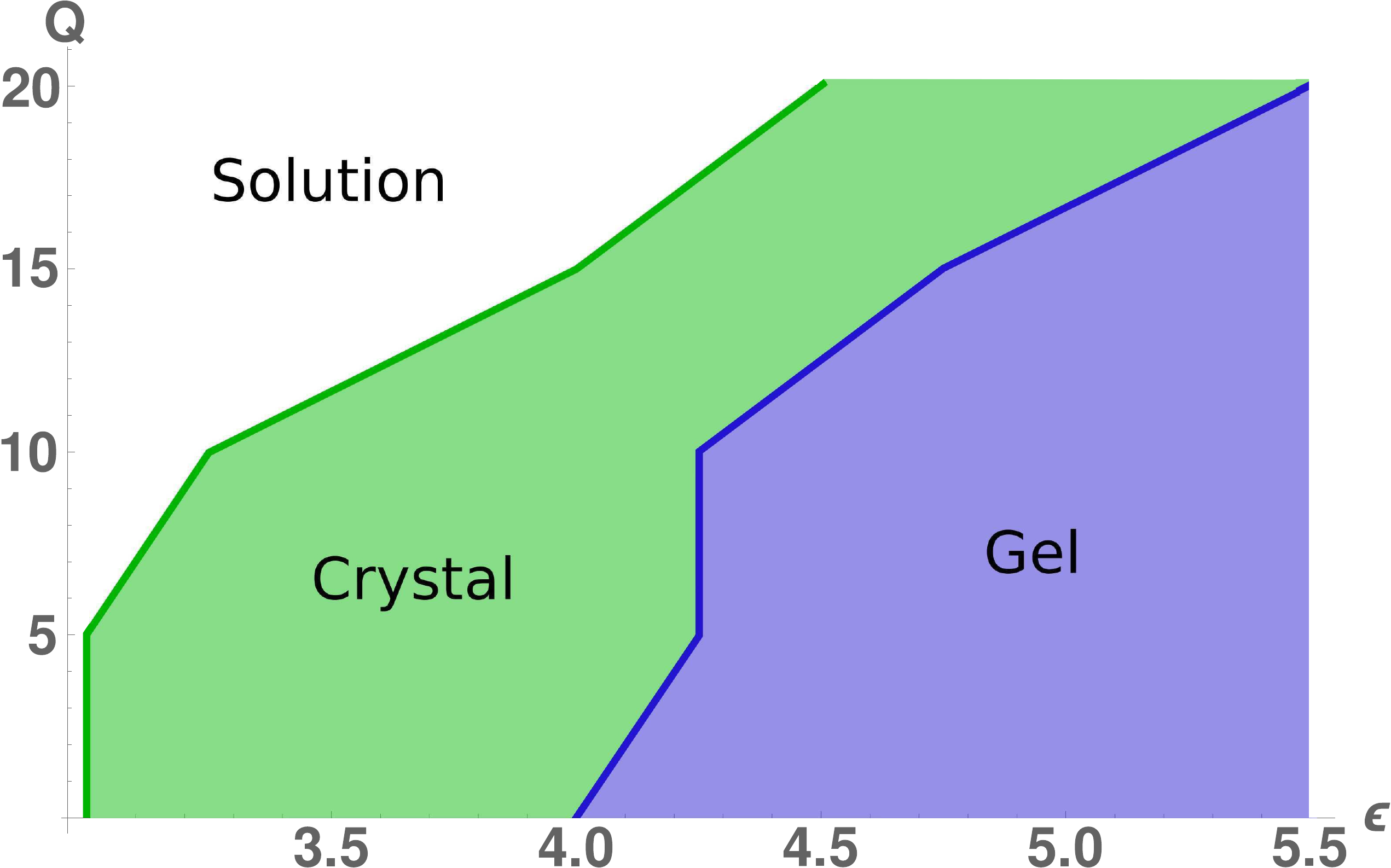}
%  \caption{Debye-H{\"u}ckel}
%  \label{fig_my_phase_dh}
%\end{subfigure}%
%\begin{subfigure}{.45\textwidth}
%  \centering
  \includegraphics[width=0.45\linewidth]{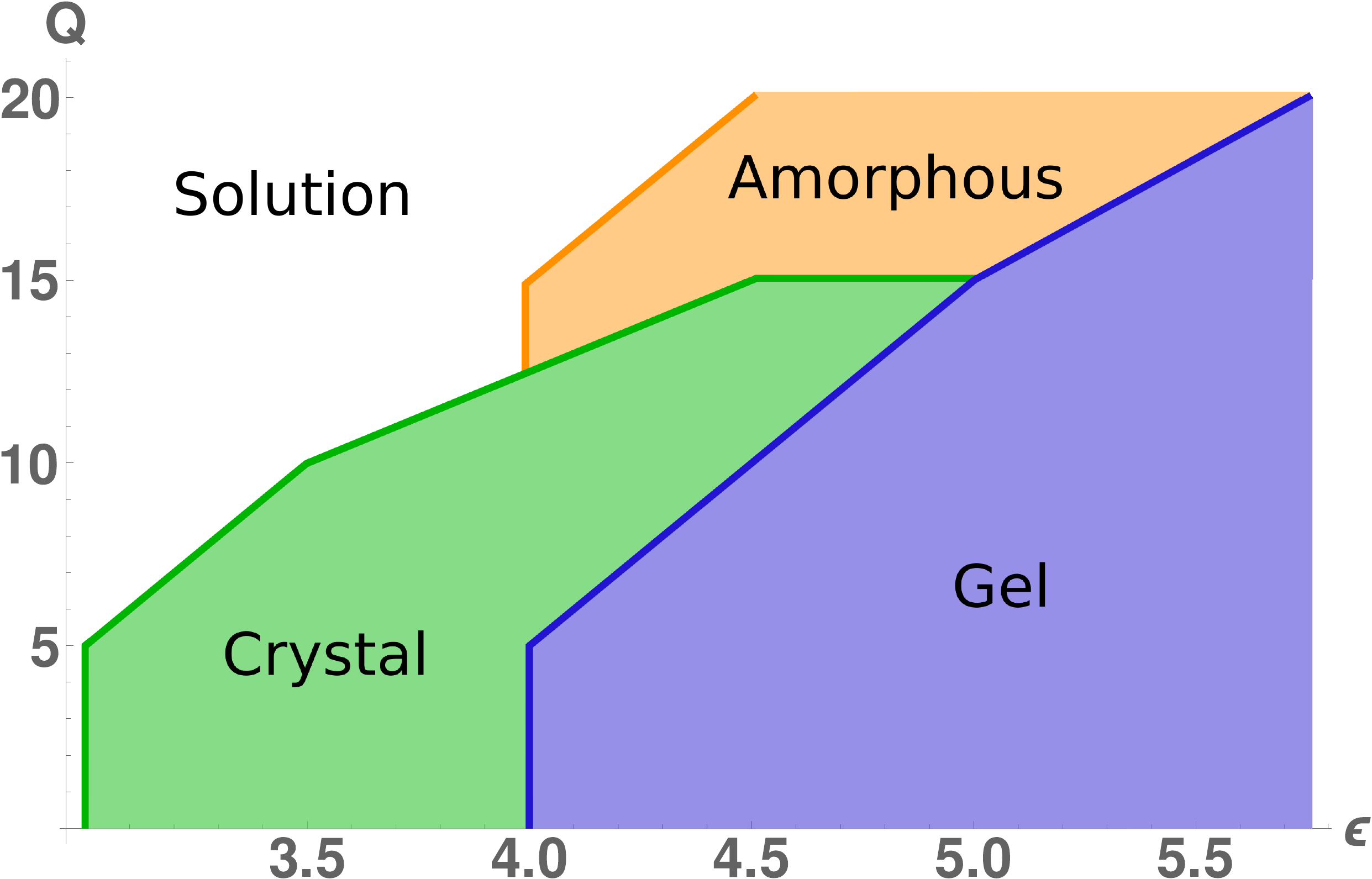}
%  \caption{Non-linear}
%  \label{fig_my_phase_nl}
%\end{subfigure}
\caption{Phase diagram for particles interacting by pairwise (left)
and many-body (right) repulsive electrostatic potentials. The net
interparticle attraction is strongest in the lower right corners
where the LJ interaction parameter $\epsilon$ is large and the
particle charge $Q$ is small. Both systems show a range of
interaction strengths favorable for crystal formation, however, this
window is truncated in the non-pairwise system by the appearance of
the amorphous phase. This phase exists because many-body effects
lead to an enhancement of the electrostatic repulsion at high
densities.} \label{fig_my_phase_nice}
\end{figure}

The left panel of Fig. \ref{fig_my_phase_nice} shows the phase
diagram for pairwise interacting particles. The net interaction
strength is weakest in the upper left (high charge, small LJ energy)
and becomes stronger moving down and to the right. This leads to the
progression from liquid, to crystal, to gel as the interaction
increases in strength. The phase diagram for particles interacting
by a non-pairwise potential is shown in the right panel of Fig.
\ref{fig_my_phase_nice}. It is similar in appearance to the phase
diagram for pairwise potentials except for the appearance of a
fourth phase in the upper right corner. This phase, which we refer
to as the amorphous phase, replaces the crystal phase in the highly
charged region of the phase diagram.

Like the gel phase, the amorphous phase spans the simulation box,
has a high surface area, and is visually opaque (Fig.
\ref{fig_pymol_all}). Based on the coordination number criteria
described above, is would be classified as a gel. However,
inspection of the coordination number histograms (Fig.
\ref{fig_phase_hist}) reveals that the amorphous phase differs
dramatically in the number of 11- and 12-fold coordinated particles,
suggesting a less densely packed structure.

\begin{figure}[!h]
%\begin{subfigure}{.45\textwidth}
%  \centering
  \includegraphics[width=0.45\linewidth]{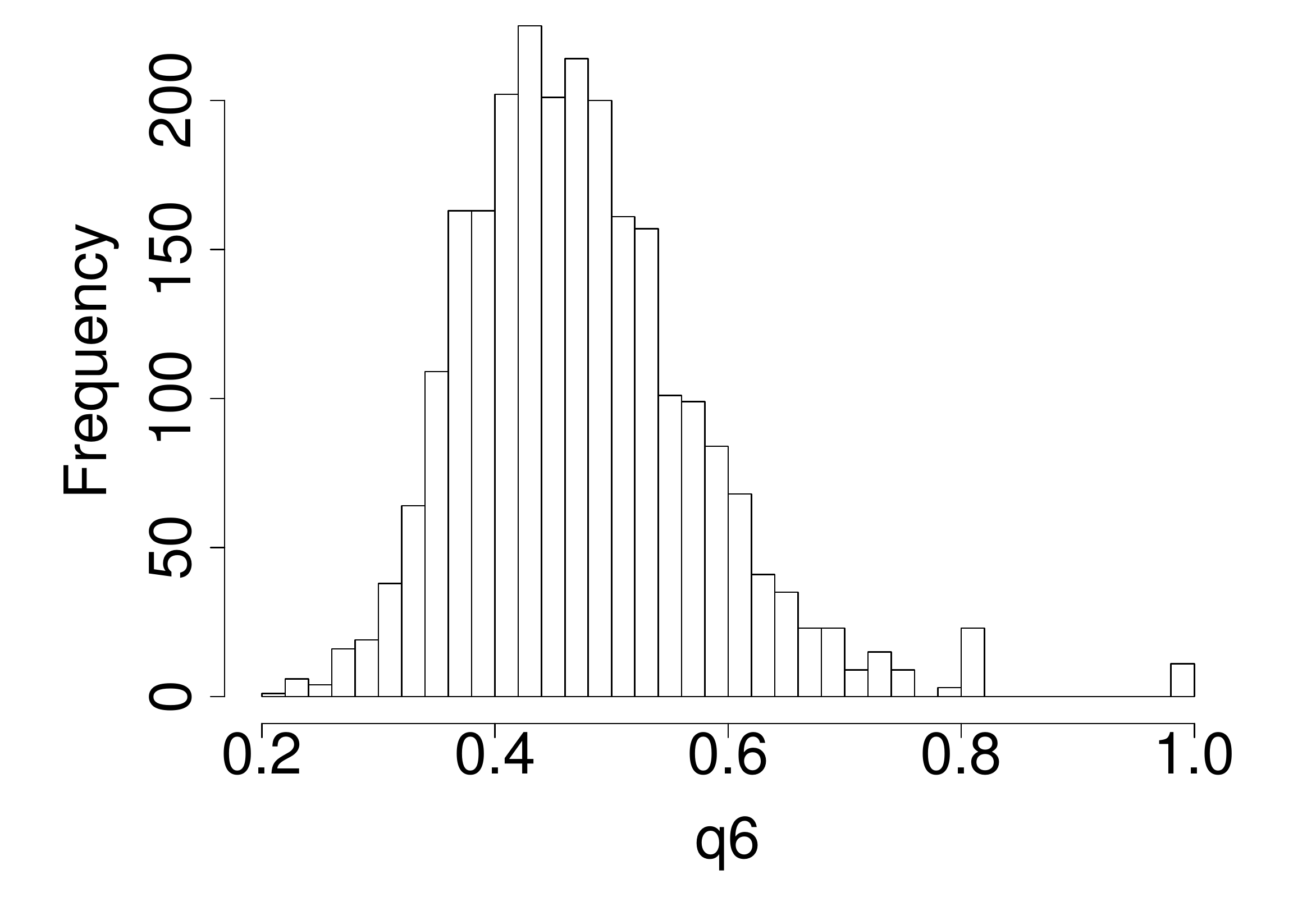}
%  \caption{Amorphous}
%  \label{fig_phase_harm_amo}
%\end{subfigure}%
%\begin{subfigure}{.45\textwidth}
%  \centering
  \includegraphics[width=0.45\linewidth]{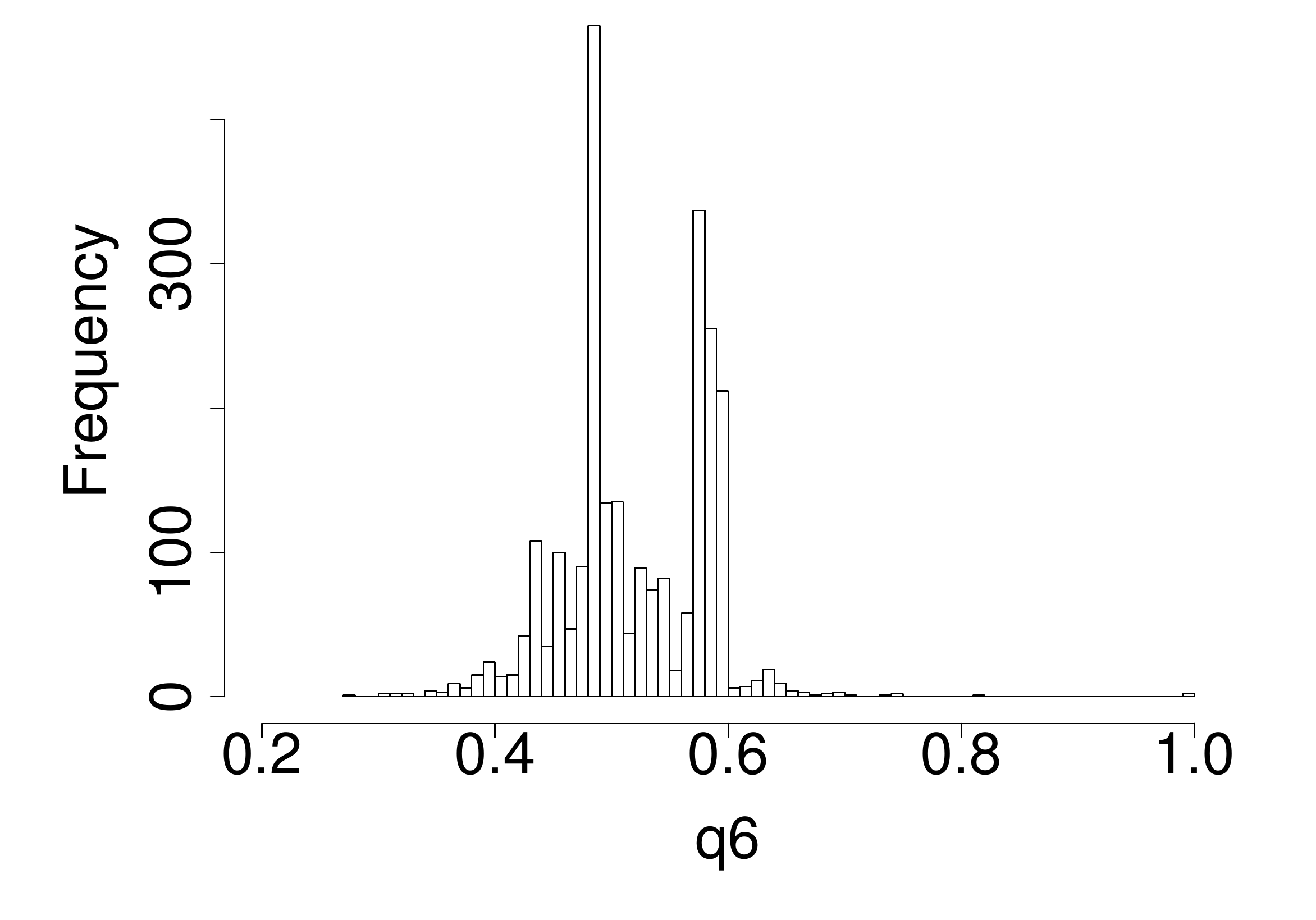}
%  \caption{Gel}
 % \label{fig_phase_harm_norm}
%\end{subfigure}
\caption{Example of an amorphous structure showing no defined fcc
($q_6=0.57$) or hcp ($q_6=0.48$) indicators in local bond order, and
a gel aggregate displaying both fcc and hcp indicator spikes.}
\label{fig_phase_harm}
\end{figure}

To further distinguish the gel and amorphous phases, we employed the
structural order parameter ${q}_l$ (Eq. \ref{eq_ql}), which projects
the position of nearest-neighbor particles onto the spherical
harmonic functions $Y_{lm}$ ~\cite{Steinhardt1983,Lechner2008}. FCC
and HCP phases are readily identified by well defined peaks at $q_6
= 0.57$ and $q_6 = 0.48$, respectively. These peaks are prominent in
the gel and crystal phases (Fig. \ref{fig_phase_harm}). In contrast,
the amorphous phase lacks defined structure in $q_6$, with a single
broad peak spanning the range 0.2 to 0.8.

%\begin{figure}[!h]
%
%    \centering
%    \includegraphics[width=1\linewidth]{qqplot}
%
%    \caption{Averaged local bond order parameters $\bar{q_6}$ and $\bar{q_4}$ for charge $Q=15$. Ordered structures roughly correspond to the region of $0.3 < \bar{Q}_6 < 0.6$ and $0 < \bar{Q}_4 < 0.2$ {\bf we need a lot more info about these plots, what conditions where they run under, what phases are we seeing, and what is the overall message that the reader should get from this?}}
%
%    \label{fig_charge15_qq}
%\end{figure}

The appearance of the amorphous phase is in qualitative agreement
with our prediction that the window of conditions favorable to
crystallization would narrow as the strength of the short-range
attraction increases \cite{Schmit2011gel}. This is because these
systems require more electrostatic repulsion to tune the net
interaction into the crystallization window, and these highly
charged particles are more strongly affected by the many-body
enhancement to the repulsion. For particles with sufficiently large
charge, the crystal phase is no longer stable. In the simple model
considered in Ref.~\cite{Schmit2011gel}, this meant that the system
remained in the soluble state. Our simulations reveal that the
system can also form the amorphous phase as a compromise. This state
allows the formation of favorable LJ interactions, but retains a low
enough density to prevent the many-body repulsion from overwhelming
the short range attraction. The extra space required to achieve this
balance prevents the system from achieving FCC, BCC, or HCP order
and allows the particles to retain a liquid-like state.

\section{Conclusion}

Control over self-assembly requires a balance between attractive and
repulsive forces. While electrostatic interactions are easily
adjustable via solution conditions, the nonlinear effects of salt
screening can be difficult to account for. Our simulations show that
under conditions where the particles are weakly charged, a pairwise
approximation can provide a good guide to system behavior. However,
more highly charged systems can differ qualitatively in their
behavior. These results suggest that special care needs to be taken
in using dilute solution properties, like the second virial
coefficient \cite{George1994,George1997}, to predict the formation
of compact states.

\begin{acknowledgments}
This work was supported by NIH Grant R01GM107487.
\end{acknowledgments}

%\bibliographystyle{h-physrev3}
%\bibliography{O:/linux/library}
%\bibliography{C:/Users/Jeremy/Documents/Work/References/library}

%merlin.mbs aipnum4-1.bst 2010-07-25 4.21a (PWD, AO, DPC) hacked
%Control: key (0)
%Control: author (8) initials jnrlst
%Control: editor formatted (1) identically to author
%Control: production of article title (-1) disabled
%Control: page (0) single
%Control: year (1) truncated
%Control: production of eprint (0) enabled
%

\end{document}